\documentclass[prl,twocolumn,showpacs,amssymb,amsfonts,amsmath,groupedaddress,floatfix,nofootinbib]{revtex4}

\usepackage{epsf,latexsym,graphicx,graphics,psfrag}
\usepackage{times}
\usepackage{amsfonts,epsfig}

\newcommand{\bra}[1]{\langle #1 |}
\newcommand{\ket}[1]{| #1 \rangle}

\newcommand{\be}{\begin{equation}}
\newcommand{\ee}{\end{equation}}
\newcommand{\ba}{\begin{eqnarray}}
\newcommand{\ea}{\end{eqnarray}}

\newcommand{\ignore}[1]{}

\def\CC{{\rm\kern.24em \vrule width.04em height1.46ex depth-.07ex
    \kern-.30em C}}
\def\P{{\rm I\kern-.25em P}}

\def\RR{{\rm
         \vrule width.04em height1.58ex depth-.0ex
         \kern-.04em R}}

\def\bbbc{{\mathchoice {\setbox0=\hbox{$\displaystyle\rm C$}\hbox{\hbox
to0pt{\kern0.4\wd0\vrule height0.9\ht0\hss}\box0}}
{\setbox0=\hbox{$\textstyle\rm C$}\hbox{\hbox
to0pt{\kern0.4\wd0\vrule height0.9\ht0\hss}\box0}}
{\setbox0=\hbox{$\scriptstyle\rm C$}\hbox{\hbox
to0pt{\kern0.4\wd0\vrule height0.9\ht0\hss}\box0}}
{\setbox0=\hbox{$\scriptscriptstyle\rm C$}\hbox{\hbox
to0pt{\kern0.4\wd0\vrule height0.9\ht0\hss}\box0}}}}

\def\bbbz{{\mathchoice {\hbox{$\sf\textstyle Z\kern-0.4em Z$}}
{\hbox{$\sf\textstyle Z\kern-0.4em Z$}} {\hbox{$\sf\scriptstyle
Z\kern-0.3em Z$}} {\hbox{$\sf\scriptscriptstyle Z\kern-0.2em
Z$}}}}

\begin{document}
\title{Ground state overlap and quantum phase transitions}
\author{Paolo Zanardi}
\author{Nikola Paunkovi\'c\footnote{Current address: SQIG, Instituto de Telecomunica\c{c}\~oes and Instituto Superior T\'ecnico, P-1049-001 Lisbon, Portugal .}}
\affiliation{ Institute for Scientific Interchange (ISI), Villa Gualino, Viale Settimio Severo 65, I-10133
Torino, Italy }
\begin{abstract}
We present a characterization of quantum phase transitions in terms of the the overlap  function between two
ground states obtained for two different values of external parameters. On the examples of the Dicke and $XY$
models, we show that the regions of criticality of a system are marked by the extremal points of the overlap and
functions closely related to it. Further, we discuss the connections between this approach and the Anderson
orthogonality catastrophe as well as with the dynamical study of the Loschmidt echo for critical systems.

\end{abstract}
\pacs{} \maketitle

\section{Introduction} Quantum phase transitions (QPT)
\cite{sachdev} have drawn a considerable interest within various
fields of physics in the recent years. They are studied in
condensed matter physics because they provide valuable information
about the novel type of finite-temperature states of matter that
emerge in the vicinity of QPT \cite{sachdev}. 
Unlike the ordinary phase transitions, driven by thermal fluctuations, QPT occur at zero temperature and are
driven by purely quantum fluctuations. In the parameter space, the  points of non-analyticity of the ground
state energy density are referred to as critical points and define the QPT. In these points one typically
witnesses the divergence of the length associated to the two-point correlation function of some relevant quantum
field. An alternative way of characterizing QPT is by the vanishing, in the thermodynamical limit, of the energy
gap between the ground and the first excited state in the critical points.
Recently, a huge interest was raised in the attempt of characterizating  QPT in terms of the notions and tools
of quantum information \cite{qis}. More specifically QPT have been studied by analyzing scaling, asymptotic
behavior and extremal points of various entanglement measures \cite{osterloh, vidal, zanardi, others,lidar}.
More recently, the connection between geometric Berry phases and QPT in
the case of the $XY$ model has been also studied \cite{carollo}.

In this paper, we aim to provide yet another characterization of
the regions of criticality that define QPT. We shall show how
 critical points can be individuated by studying a surprisingly simple
quantity: the overlap i.e., the scalar product, between two ground states corresponding to two slightly
different values of the parameters. The physical intuition behind this approach should be obvious: QPT mark the
separation between regions of the parameter space which correspond to ground states having deeply different
structural properties e.g., order parameters. This difference is here quantified by the simplest Hilbert-space
geometrical quantity i.e., the overlap. Note that the square modulus of the overlap is nothing but the fidelity,
widely used in quantum information as a function that provides the criterion for distinguishability between
quantum states \cite{qis}. Therefore, it is a natural candidate for a study of macroscopic distinguishability
between quantum states that define different macroscopic states of matter (different phases). When applied to
cases of many-body systems containing many degrees of freedom, the overlap (or, fidelity) might seem to be too
coarse quantity, and not bearing any apparent information about the difference in order properties between
quantum phases, to be of any use. Nevertheless the main result of this paper  is that in some cases it is indeed
possible to do so.
The critical behavior of a system undergoing QPT is reflected in the geometry of its Hilbert space: approaching
the QPT the overlap (distance) between neighboring ground states shows a dramatic drop (increase). We would like
to notice that Cejnar {\em et. al.} \cite{cejnar} discussed the overlap entropy between the eigen-bases of a
system's Hamiltonian and various physically relevant bases, in the context of enhanced decoherence effects in
the regions of criticality (see also \cite{other}).


In the following two sections, we conduct our analysis on the
cases of two simple, yet physically relevant and mathematically
instructive, examples of the Dicke model and the $XY$
spin-chain model. Next, we discuss the connection between the
scaling and asymptotic behaviors and the so-called Anderson
orthogonality catastrophe \cite{anderson}. Moreover, the relation
with the  dynamical study of decoherence and quantum criticality
 \cite{zanardi-china} is briefly addressed.
Finally, in the last section conclusions are discussed.


For a generic point in parameter space we use label $q\in\mathbb{R}^L$, where $L$ is the number of external
parameters determining system's Hamiltonian. As the overlap function depends on the difference between
parameters as well, we introduce $\tilde{q} \equiv q + \delta q$ to denote the neighboring point $\tilde{q}$ and
the difference $\delta q$. Following this notation, we denote the ground states by $\ket{g} \equiv \ket{g(q)}$
and $\ket{\tilde{q}} \equiv \ket{g(\tilde{q})}$. In general, all the functions $F(q)$ evaluated in the point
$\tilde{q}$ we will denote as $\tilde{F}$, while those evaluated in the critical point $q_c$, we will denote as
$F_c$ (note that by combining two cases, we have $\tilde{F}_c=F(q_c+\delta q)$). Then, the overlap function is
simply given by the scalar product $\langle g(q) | g(\tilde{q}) \rangle$ (note that all the results of this
paper could be easily formulated in terms of fidelity as well).
We shall examine the behavior of
the overlap  as a function of $q$ only, while keeping
$\delta q$ fixed and small.

\section{ The Dicke Model} Our first example is the Dicke
model. It describes a dipole interaction between a
single bosonic mode $\hat{a}$ and a collection of $N$ two-level
atoms. If for $N$ atoms we introduce the collective angular
momentum operators $\hat{J}_s, s\in\{\pm,z\}$, Dicke Hamiltonian
has the following form (we take $\hbar=1$): \be \label{Dicke
Hamiltonian} \hat{H}(\lambda) = \omega_0\hat{J}_z +
\omega\hat{a}^\dag\hat{a} +
\frac{\lambda}{\sqrt{2j}}\left(\hat{a}^\dag+\hat{a}\right)\left(\hat{J}_+
+ \hat{J}_-\right). \ee Parameter $\lambda$ is the atom-field
coupling strength and is the one driving the QPT in this model.
Therefore, we have $q=\lambda$ and denote the Hamiltonian's
dependance on that parameter only. Parameters $\omega_0$ and
$\omega$ stand for the atomic level-splitting and bosonic mode
frequency, respectively, while $j$ describes the length of a
collective spin vector, and is assumed to be constant and equal to
$j=N/2$. In the thermodynamical limit $(N\rightarrow \infty)$,
which is here equivalent to $(j\rightarrow \infty)$, Dicke
Hamiltonian undergoes a quantum phase transition for the critical
value of its parameter $\lambda$ given by
$\lambda_c=(\omega\omega_0)/2$. When $\lambda<\lambda_c$, the
system is in highly unexcited {\em normal} phase, while
$\lambda>\lambda_c$ defines the {\em super-radiant} phase in which
both the field and $N$ atoms become macroscopically excited. The
super-radiant phase is characterized by the broken symmetry given
by the parity operator $\hat{\Pi} = \exp(i\pi\hat{N}),
\hat{N}=(\hat{a}^\dag\hat{a}+\hat{J}_z+j)$: the ground state is
doubly degenerate. As shown in \cite{emary}, by introducing
bosonic operators $\hat{b}$ through Holstein-Primakoff
representation \cite{primakoff}, the above Dicke Hamiltonian
(\ref{Dicke Hamiltonian}) can be exactly diagonalized in the
thermodynamical limit. In the normal phase, its form is: \be
\label{Dicke - Normal} \hat{H}^n(\lambda) =
\omega_0\hat{b}^\dag\hat{b} + \omega\hat{a}^\dag\hat{a} +
\lambda\left(\hat{a}^\dag+\hat{a}\right)\left(\hat{b}^\dag +
\hat{b}\right) - j\omega_0.\ee Its ground state is:
$g(x,y)=\left(\frac{\varepsilon_+\varepsilon_-}{\pi^2}\right)^{\frac{1}{4}}e^{-1/2\langle
{\bf{R}}, A{\bf{R}}\rangle} [{\bf{R}}=(x,y)]$ where $x$ and $y$
are the real space coordinates associated to the modes $\hat{a}$
and $\hat{b},$ $A=U^{-1}MU$,
$M=\text{diag}[\varepsilon_-,\varepsilon_+]$ and
$U$ an orthogonal matrix $U=\left[ \begin{array}{cc} c & -s \\
s & c
\end{array}\right]$, ($c=\cos\gamma, s=\sin\gamma$ are given by the squeezing angle
$\gamma=(1/2)\arctan[4\lambda\sqrt{\omega\omega_0}/(\omega^2+\omega_0^2)]$). $\varepsilon_{\pm}$ represent the
fundamental collective excitations of the system and are given by:
$\varepsilon^2_{\pm}=\frac{1}{2}\left(\omega^2+\omega_0^2\pm\sqrt{(\omega^2-\omega_0^2)^2+16\lambda^2\omega^2\omega_0^2}\right).$
From the above formula, we see that $\varepsilon_-(\lambda_c)\equiv\varepsilon^c_-=0$: the system becomes
gapless and undergoes a QPT for $\lambda=\lambda_c$.

The overlap,  calculated between two ground states $g$ and $\tilde{g}$, is given by
\be \label{DickeOverlap} \langle g|\tilde{g}\rangle\!=\!2\frac{[\det\! A \det\!
\tilde{A}]^{\frac{1}{4}}}{[\det(A\!\!+\!\!\tilde{A})]^\frac{1}{2}}\!\!=\!\!2\frac{[\det\!
A]^{\frac{1}{4}}}{[\det\! \tilde{A}]^\frac{1}{4}[\det(1\!\!+\!\!\tilde{A}^{-1}\!\!A)]^\frac{1}{2}}.\ee Note that
the overlap  is a function of both $\lambda$ and $\delta\lambda$. In the limit
$(\lambda\rightarrow\lambda_{c})$, with $\delta\lambda>0$ being fixed, $\det
A=\varepsilon_+\varepsilon_-\rightarrow 0$, while $\det \tilde{A}\geq\det
\tilde{A}_c=\tilde{\varepsilon}^c_+\tilde{\varepsilon}^c_->0$. The same holds for $\det(1+\tilde{A}^{-1}A)$, for
a sufficiently small $\delta\lambda$ (note that $\lim_{\delta\lambda\rightarrow 0}\tilde{A}^{-1}=A^{-1}$). But
in the present case, it is possible to show that for every, and not just small $\delta\lambda$,
$\det(1+\tilde{A}^{-1}A)$ does not vanish. Using the formula $\det (1+A)=1+\mbox{Tr}A+\det A$ for $2\times 2$
matrices, we get: $\det(1+\tilde{A}^{-1}A)\rightarrow 1+\mbox{Tr}(\tilde{A}_c^{-1}A_c)$ (note that
$\det(\tilde{A}^{-1}A)=[\det\tilde{A}]^{-1}\det A\rightarrow 0$). After a straightforward calculation, we obtain
the result: $
\mbox{Tr}(\tilde{A}_c^{-1}A_c)=\frac{\varepsilon^c_+}{\tilde{\varepsilon}^c_+\tilde{\varepsilon}^c_-}\left[
(s\tilde{c}+c\tilde{s})^2\tilde{\varepsilon}^c_+ + (s\tilde{s}-c\tilde{c})^2\tilde{\varepsilon}^c_-\right].$
Therefore, $\mbox{Tr}(\tilde{A}_c^{-1}A_c)>0$ for every $\lambda$ and we can conclude that $\langle
g|\tilde{g}\rangle\propto (\varepsilon_-)^{1/4}$ as $(\lambda\rightarrow\lambda_c)$. In Ref.~\cite{emary}, it
was shown that when approaching the critical point from both normal and super-radiant sides, the excitation
energy $\varepsilon_-$ drops as the square root of $\Delta\equiv |\lambda-\lambda_c|$, which gives us the
asymptotic behavior of the overlap  function in the vicinity of the critical point: $\langle
g|\tilde{g}\rangle\propto \Delta^{1/8}$. Although we have provided here only the results for overlap function
for the system in the normal phase, the analogous analysis for the super-radiant phase gives us the same
qualitative results, as the two ground states are again the Gaussian-type states, but with translated and
re-scaled $x$ and $y$ axes. Therefore, we omit it here.



\begin{figure}[ht]
\includegraphics[width=6.5cm,height=3.7cm,angle=0]{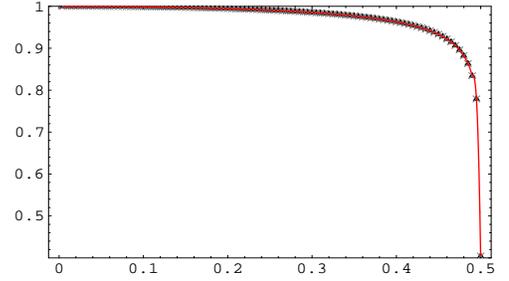}
\caption{(color online) The overlap  function $\langle g |
\tilde{g}\rangle$, equation (\ref{DickeOverlap}), as a function of
$\lambda<\lambda_c$, taken for the resonant case
$\omega_0=\omega=1$ and $\delta\lambda=10^{-6}$. Note the dramatic
decreasing of the function as we approach the point of
criticality.} \label{Dicke Overlap - Figure}
\end{figure}


We conclude this section by presenting the numerical results for the overlap  function in the normal phase. In
Fig. \ref{Dicke Overlap - Figure} we plot the overlap (\ref{DickeOverlap}) between two ground states of Dicke
Hamiltonian for the resonant case $\omega_0=\omega=1$ and $\delta\lambda=10^{-6}$. We see that it is almost
constant and equal to unity for wide range of $\lambda$, apart from the very narrow area around $\lambda_c$,
when it drastically drops to zero. Such behavior of the overlap  function around the point of criticality can be
ascribed to the fact that the ground state for $\lambda=\lambda_c$ becomes completely delocalized along one of
two rotated axes, as opposed to the localized ground state outside of the point of criticality (see
\cite{emary}).


\section{The XY Spin Chain} In the following section, we
discuss the example of the one-dimensional $XY$ anisotropic
spin-half chain in the external magnetic field. Its Hamiltonian is
given by the following expression:

\be \label{Hamiltonian} \hat{H}(\gamma, \lambda)\! =\!\! -\!\!\!\!\sum_{i = -M}^{M} \!\!\! \left(
\frac{1+\gamma}{2}\hat{\sigma}_{i}^x \hat{\sigma}_{i+1}^x\!\! +\!\! \frac{1-\gamma}{2} \hat{\sigma}_{i}^y
\hat{\sigma}_{i+1}^y\!\! +\!\! \frac{\lambda}{2}\hat{\sigma}_{i}^z\right). \!\!\ee The parameter $\gamma \in
\mathbb{R} $ defines the anisotropy, while $\lambda \in \mathbb{R}$ represents external magnetic field along the
$z$ axis, up to a factor $\frac{1}{2}$. Therefore, $q=(\gamma,\lambda)$. The operators
$\hat{\sigma}_{i}^{\alpha},\, \alpha \in \{ x,y,z\}$ are the usual Pauli operators. This Hamiltonian can be
exactly diagonalized by successively applying Jordan-Wigner, Furier and Bogoliubov transformation (see for
example \cite{sachdev}). This way, we obtain the following form of the Hamiltonian: $ \hat{H}(\gamma, \lambda) =
\sum _{k = -M}^M\Lambda_k(\hat{b}^{\dagger}_k \hat{b}_k - 1).$ The energies of one-particle excitations are
given by $\Lambda_k=\sqrt{\varepsilon_k^2+\gamma^2\sin^2\frac{2\pi k}{N}},$ with $\varepsilon_k=\cos\frac{2\pi
k}{N}-\lambda$ and $N=2M+1$ being the total number of sites (spins). One-particle excitations are given by the
fermionic operators $\hat{b}_k=\cos \frac{\theta_k}{2}\hat{d}_k -i\sin\frac{\theta_k}{2}\hat{d}^{\dag}_{-k},$
with $ \cos\theta_k = \varepsilon_k/\Lambda_k$.
Finally, the
ground state $\ket{g(\gamma, \lambda)}$, that is defined as the
state to be annihilated by each operator $\hat{b}_k$
($\hat{b}_k\ket{g(\gamma, \lambda)}\equiv 0$), is given as a
tensor product of qubit-like states: \be \label{ground_state}
\ket{g(\gamma, \lambda)}\!\! = \!\!\bigotimes_{ k =
1}^M\!\!\left(\!\! \cos\frac{\theta_k}{2} \ket{0}_k\ket{0}_{-k}
\!\!-i \sin\frac{\theta_k}{2}\ket{1}_k\ket{1}_{-k} \!\!\right).
\ee

In its space of parameters, the family of Hamiltonians given by
equation (\ref{Hamiltonian}) exhibits two regions of criticality,
defined by the existence of gapless excitations: {\em (i)} $XX$
region of criticality, for $\gamma=0$ and $\lambda \in (-1,1)$;
{\em (ii)} $XY$ region of criticality given by the lines $\lambda
= \pm 1$.

\begin{center}
\begin{figure}[tp]
\begin{picture}(10,0)
\put(-30,50){(a)}\put(8,50){\tiny{1}}\put(3,27){\tiny{0.9996}} \put(-30,-45){(b)}\put(-8,-35){\tiny{$3.5\times
10^9$}}\put(3,-65){\tiny{0}} \put(-30,-133){(c)}\put(-8,-127){\tiny{$2.5\times 10^8$}}\put(3,-155){\tiny{0}}
\end{picture}
\psfrag{l}[Bc][][0.75][0]{$\lambda$} \psfrag{g}[Bc][][0.75][0]{$\gamma$} \psfrag{gg}[Bc][][0.75][0]{$$}
\psfrag{l}[Bc][][0.75][0]{$\lambda$}%
\psfrag{g}[Bc][][0.75][0]{$\gamma$} \psfrag{gg}[Bc][][0.75][0]{$$}
\includegraphics[width=4.5cm,height=3.2cm,angle=0]{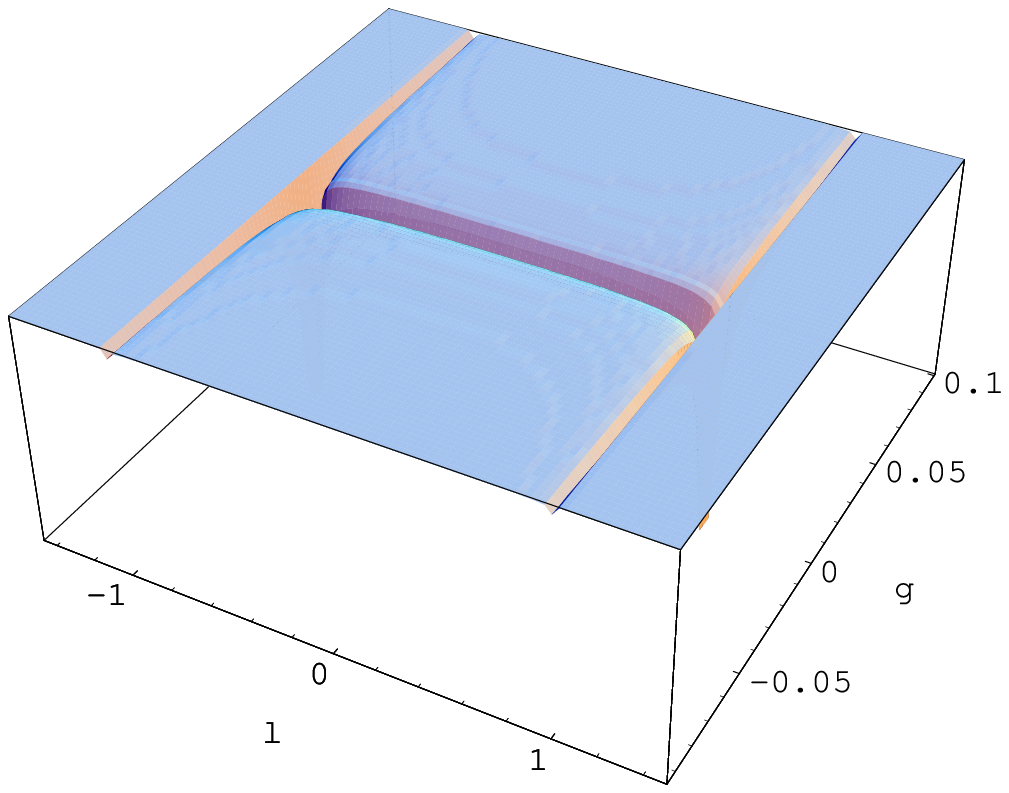}\hskip
2mm
\includegraphics[width=4.5cm,height=3.2cm,angle=0]{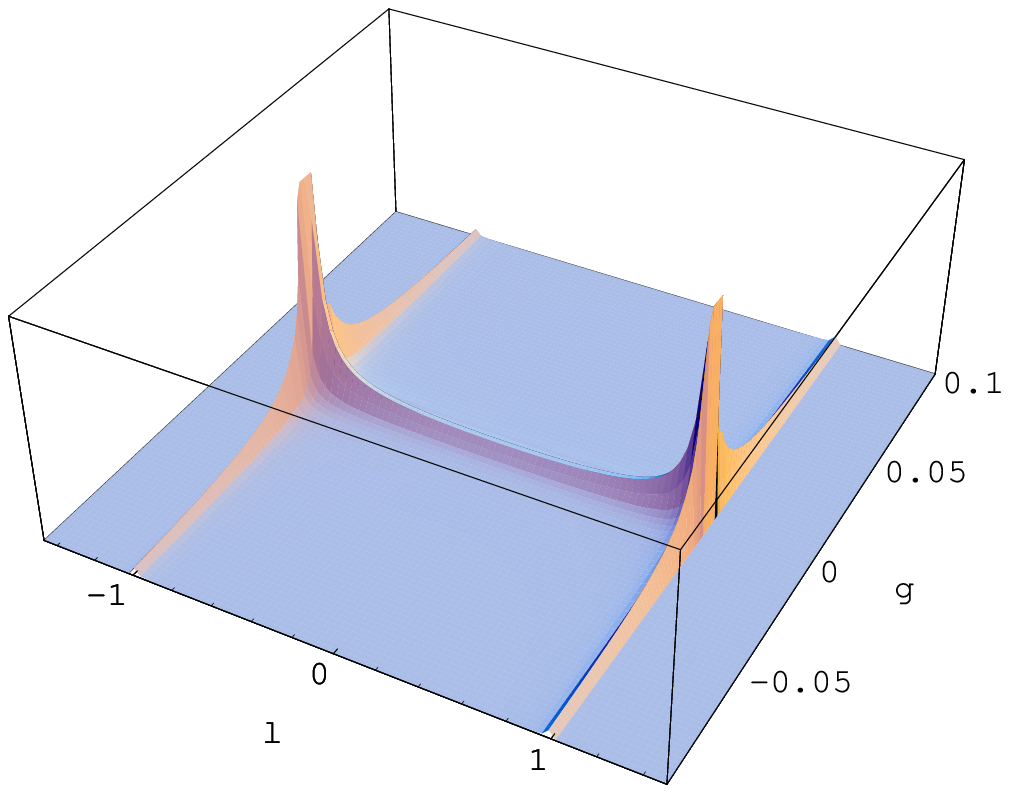}\hskip
2mm
\includegraphics[width=4.5cm,height=3.2cm,angle=0]{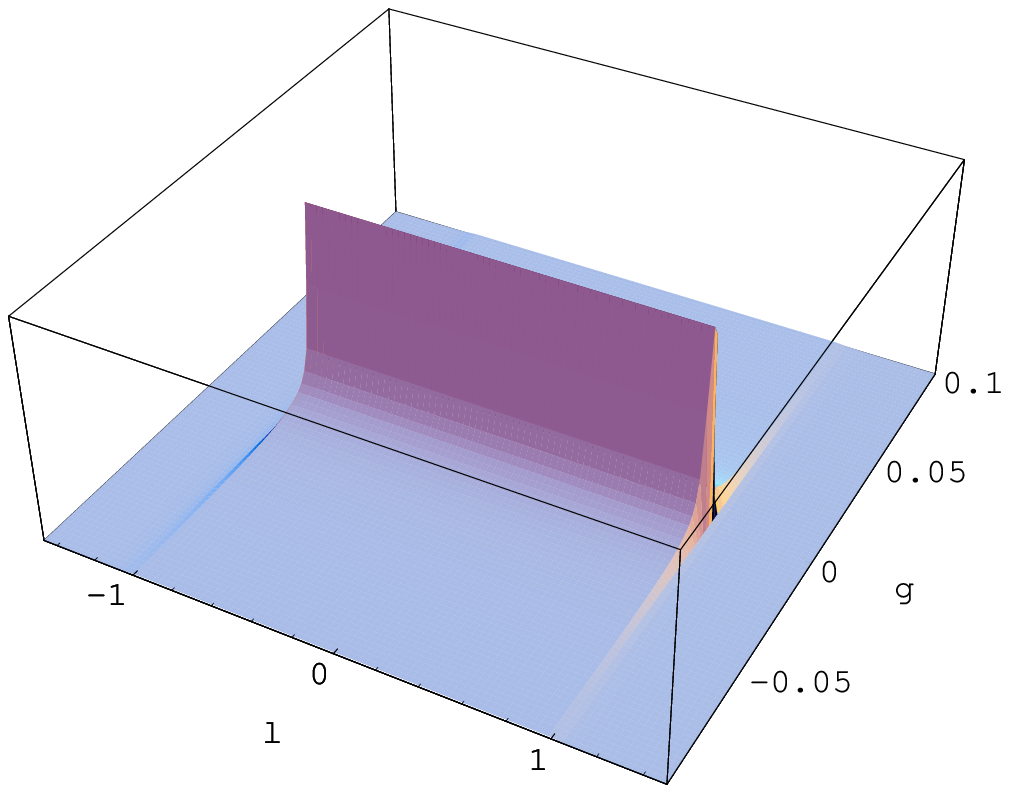}
\hskip -2mm \caption{(color online) (a) The overlap  function $\langle g(q) | g(\tilde{q})\rangle$, as a
function of $\lambda$ and $\gamma$, for $N=10^6$ and $\delta\lambda=\delta\gamma=10^{-6}$. Note the clear dips
of the plot in the regions of criticality. (b) $S^{\lambda}_N(\lambda, \gamma)$. (c) $S^{\gamma}_N(\lambda,
\gamma)$.} \label{XY-figure}
\end{figure}
\end{center}

As in the previous example, let us first consider the exact
overlap  function. From equation (\ref{ground_state}), it follows
that the exact overlap  function between the ground states
$\ket{g}$ and $\ket{\tilde{q}}$ is: \ba \label{overlap-exact}
\langle g(q) | g(\tilde{q}) \rangle = \prod_{k=1}^M
\cos\frac{\theta_k - \tilde{\theta}_k}{2}, \ea where
$\tilde{\theta}_k=\theta_k(\tilde{q})$. Note the dependence on the
number of sites $N$ that is implicit in all the previous formulae
from this section. In Fig. \ref{XY-figure}(a), we present the
numerical result obtained using the above equation
(\ref{overlap-exact}), for $N=10^6$ spins and
$\delta\lambda=\delta\gamma=10^{-6}$. We observe that the regions
of criticality are clearly marked by a sudden drop of the value of
the overlap  function. As before, we ascribe this type of behavior
to a dramatic change in the structure of the ground state of the
system while undergoing QPT.

In order to investigate the overlap  function more quantitatively
and relate its behavior to the existence of the regions of
criticality, we note that while the overlap depends on the values
of both the parameters $q$ and the difference $\delta q$, the
regions of criticality are defined by the values of parameters
only. Therefore, in the following we choose to study the functions
\be \label{Sl} S^{\lambda}_N(\lambda, \gamma) \equiv
\sum_{k=1}^M\left(\frac{\partial\theta_k}{\partial\lambda}\right)^2,
S^{\gamma}_N(\lambda, \gamma) \equiv
\sum_{k=1}^M\left(\frac{\partial\theta_k}{\partial\gamma}\right)^2,\ee
that define the first non-zero order of the Taylor expansion of
the overlap  function (\ref{overlap-exact}).
Functions $S^{\lambda}_N(\lambda, \gamma)$ and
$S^{\gamma}_N(\lambda, \gamma)$ are natural candidates for our
study because they express the ``rate of change" of the ground
state, taken in the point $q$. They do not depend on the
difference $\delta q$, and although for every finite $N$ it is
possible to find $\delta q$ small enough so that the exact overlap
is arbitrarily well approximated by the expression $\exp
(-\frac{1}{8}S^q_N(q)\delta q^2)$, functions
$S^{\lambda}_N(\lambda, \gamma)$ and $S^{\gamma}_N(\lambda,
\gamma)$ on their own capture the behavior of the $\langle g(q) |
g(\tilde{q}) \rangle$ function and are enough for our current
study. They also allow for analytic investigation, together with
numerical one. In Figs. \ref{XY-figure}(b) and \ref{XY-figure}(c)
we present the numerical results for $S^{\lambda}_N(\lambda,
\gamma)$ and $S^{\gamma}_N(\lambda, \gamma)$, respectively, for
$N=10^6$ spins. Again, the regions of criticality could easily be
inferred by simply observing both plots. Note that in this case
the relative difference between the numerical values in the
regions of criticality and elsewhere is much
bigger than in the case of the exact overlap (see Fig.
\ref{XY-figure}(a)). But now, {\em both} plots are needed to
detect both regions of criticality. This is so because by moving
along $\gamma=0$, while keeping $|\lambda|<1$, we do not move
outside the $XX$ region of criticality and therefore do not expect
the qualitative change in the structure of the ground state, and
consequently in the behavior of $S^{\lambda}_N(\lambda, \gamma)$
as well. The same holds for $S^{\gamma}_N(\lambda, \gamma)$ and
the $XY$ region of criticality.


We first examine the scaling behavior of $S^{\lambda}_N(\lambda, \gamma)$ and $S^{\gamma}_N(\lambda, \gamma)$
with respect to number of spins $N$. The numerics present us with the following results. First, as expected,
$S^{\lambda}_N(\lambda, \gamma)$ and $S^{\gamma}_N(\lambda, \gamma)$ scale linearly with $N$ when
$(\gamma\rightarrow 0)$ and $(\lambda\rightarrow \pm 1)$, respectively. In the regions of criticality, we have
that $S^{\lambda}_N(|\lambda|=1, \gamma)\propto N^2/\gamma^2$, while on the other hand, for $(\gamma\rightarrow
0)$ we still have $S^{\gamma}_N(\lambda, \gamma)\propto N$, without being a function of $\lambda$ (we note here
that $S^{\gamma}_N(|\lambda|<1, \gamma=0)=0$). Such different behavior is a consequence of the fact that while
the $XY$ region of criticality defines the second order QPT, $XX$ is the example of the third order QPT.

We have also conducted a separate analytical study, confirming the above numerical results. First, we note that
for every point $q$ in parameter space, and every {\em finite} $N$, partial derivatives
$\left(\frac{\partial\theta_k}{\partial\lambda}\right)$ and
$\left(\frac{\partial\theta_k}{\partial\gamma}\right)$  are continuous functions of the parameters. They can
become infinite only in the thermodynamical limit, when $(N \rightarrow \infty)$, and only in the {\em regions
of criticality}. By looking at the explicit form of derivatives (we use $x_k = \frac{2\pi}{N}k$):\be \label{Dl}
\left(\frac{\partial\theta_k}{\partial\lambda}\right) = \frac{\gamma(\sin x_k)}{[(\cos x_k - \lambda)^2 +
\gamma^2(\sin x_k)^2]},\ee

\be \label{Dg} \left(\frac{\partial\theta_k}{\partial\gamma}\right) =
 - \frac{|\sin x_k|(\cos x_k - \lambda)}{[(\cos x_k - \lambda)^2 +
\gamma^2(\sin x_k)^2]},\ee we see that only when the energy $\Lambda_k$ (the denominator of both of the above
expressions) gets arbitrarily small (or zero), the derivatives (\ref{Dl}) and (\ref{Dg}) can become divergent.
In other words, only when $\cos x_k$ gets arbitrarily close to $\lambda$, {\em and} either $\gamma$ or $\sin
x_k$ get close to zero. That is, in the regions of criticality. Note that we assume that for every $N \in
\mathbb{N}$ and $k \in \{ 1,\ldots M \}$, equation $\cos x_k = \lambda$ has no solution, which presents a
generic case ($\lambda$'s that allow for the solutions of this equation form a set of measure zero on the
$(-1,1)$ interval). Therefore, outside the regions of criticality $S^{\lambda}_N(\lambda, \gamma)$ and
$S^{\gamma}_N(\lambda, \gamma)$ scale linearly with $N$.

Regarding the regions of criticality, we  first consider the scaling behavior of $S^{\lambda}_N(\lambda,
\gamma)$ in the vicinity of the $XX$ criticality. As there always exists $k_0$ such that in the $(N\rightarrow
\infty)$ limit $\cos x_{k_0} \rightarrow \lambda$, then for such $x_{k_0}$ and every finite $\gamma$, it follows
from (\ref{Dl}) that $\left(\frac{\partial\theta_{k_0}}{\partial\lambda}\right) \rightarrow (\gamma \sin
x_{k_0})^{-1}$, when $(N\rightarrow \infty)$. In other words, it does not scale with $N$ (note that although
$k_0=k_0(N)$ is a function of $N$, $\lim_{N\rightarrow\infty}\sin x_{k_0}=\sin\arccos\lambda$). As all other
derivatives are finite, we have that $S^{\lambda}_N(|\lambda|<1, \gamma\rightarrow 0)\propto N/\gamma^2$.
Similar discussion can be applied to the case of $S^{\gamma}_N(\lambda, \gamma)$ in the $XY$ region of
criticality. Again, there exists a qubit defined by $k_1=1$ for which $\cos x_{k_1} \rightarrow 1$ in the
$(N\rightarrow \infty)$ limit, so that its existence could bring about the scaling of $S^{\gamma}_N(\lambda,
\gamma)$ larger than linear in thermodynamical limit. Using the Taylor expansion of sine and cosine functions
around zero (note that in that case, $\sin x_{k_1}\rightarrow0$ as well), from equation (\ref{Dg}) we obtain
$\left(\frac{\partial\theta_k}{\partial\gamma}\right)\propto x_{k_1}/\gamma^2\rightarrow 0,
(N\rightarrow\infty)$. In other words, $S^{\gamma}_N(|\lambda|=1, \gamma)\propto N$.

Now, we turn to more interesting cases of the relevant functions $S^{\lambda}_N(\lambda, \gamma)$ and
$S^{\gamma}_N(\lambda, \gamma)$, in the $XY$ and $XX$ regions of criticality, respectively. Using Taylor
expansions of sine and cosine functions around zero, we see that in $\lambda = \pm 1$ the derivative
$\left(\frac{\partial\theta_{k_1}}{\partial\lambda}\right)$ given by $k_1=1$ behaves like
$\left(\frac{\partial\theta_{k_1}}{\partial\lambda}\right) \propto N/(2\pi\gamma)$ as $(N\rightarrow \infty)$
(see equation (\ref{Dl})) and therefore $S^{\lambda}_N(|\lambda|=1, \gamma)\propto N^2/\gamma^2$. We also see
that the scaling factor depends on $\gamma$. Finally, from (\ref{Dg}), we also see that $S^{\gamma}_N(\lambda,
\gamma)\propto N$.

The alternative way to examine the signatures of QPT is to look at the asymptotic behavior of two functions
(\ref{Sl}) near the regions of criticality. From the numerical study we obtain that the asymptotic behavior of
$S^{\lambda}_N(\lambda, \gamma)$ in the vicinity of critical points $\lambda_c = \pm1$, $\gamma \in (0,1]$, is
given by the following formula: $S^{\lambda}_N(\lambda, \gamma)\propto
a(\gamma,N)/|1-\lambda|^{\alpha(\gamma,N)}$. From the study of the scaling behavior, we already know that
$a(\gamma,N)=a(\gamma)N^2$ and that $a(\gamma) \propto 1/\gamma^2$. Further, from numerics we have that the
exponent $\alpha(\gamma,N)$ is constant with respect to $\gamma$ and approaches to $\alpha=1$ as $(N\rightarrow
\infty)$. Such asymptotic behavior, with constant exponent $\alpha=1$ for all $\gamma \in (0,1]$ could be seen
as a consequence of the fact that in that range of parameters the $XY$ model belongs to the same class of
universality. The numerics gives also the asymptotic behavior of $S^{\gamma}_N(\lambda, \gamma)$ in the vicinity
of $\gamma=0$ (with $|\lambda|<1$) similar to the previous one, $ S^{\gamma}_N(\lambda, \gamma)\propto
b(\lambda,N)/\gamma^{\beta(\lambda,N)}$, with the exponent $\beta(\lambda,N)$ approaching to $\beta=1$ as
$(N\rightarrow \infty)$. But, the coefficient $b(\lambda,N)$ depends only on $N$, and as noted before, scales
linearly with it, $b(\lambda,N)\propto bN$.

\section{QPT: Orthogonality catastrophe, Loschmidt echo} The above two examples represent a generic case of a
many-body system which exhibits continuous QPT only in the thermodynamical limit.
In the case of the $XY$ model, as the number of spins increases, the overlap between two different ground states
(\ref{ground_state}) approaches to zero, no matter how small the difference in parameters $\delta q$ is, so that
in thermodynamical limit each two ground states are mutually orthogonal;  they live in a continuous tensor
product space \cite{ITP}.
Such behavior of systems having infinitely many degrees of
freedom, when the two physical states corresponding to two
arbitrarily close sets of parameters (two arbitrarily similar
physical situations) become orthogonal to each other, has been
already studied in many-body physics and is known as Anderson {\em
orthogonality catastrophe} \cite{anderson}. From our study of the
$XY$ model, we have seen that not only that every two ground
states become orthogonal in thermodynamical limit, but also the
rate of ``orthogonalization" between two ground states of large,
but finite system, changes qualitatively and grows faster in the
vicinity of the regions of criticality.
This way, the regions of criticality of an infinite system are
already marked by the scaling and asymptotic behavior of the
relevant functions of a finite-size system. Loosely speaking, the
regions of criticality of QPT are given as regions where the
orthogonality catastrophe is expressed on qualitatively greater
scale. Notice that recently, the occurrence of a particular
instance of Anderson-type orthogonality catastrophe was studied
for the case of a system in the vicinity of QPT
\cite{sachdev-catastrophe}.

Now we would like to establish an explicit  connection between the
sort of kinematical approach used in this paper and the dynamical
one of Ref. \cite{zanardi-china}. In order to do so let us
introduce the projected density of states function
$D(\omega;q,\tilde{q}) \equiv \bra{g(\tilde{q})}\delta(\omega -
\hat{H}(q))\ket{g(\tilde{q})}$ that describes the spread of the
ground state $\ket{g(\tilde{q})}$ expressed in the eigenbasis
obtained for the point $q$. Then, the square of the overlap  can be expressed as
\begin{equation}
|\langle g(q)|g(\tilde{q})\rangle |^2= 1 -
\int_{E_1}^{\infty}D(\omega)d\omega
\end{equation}
 ($E_1$ denotes the first
excited eigenvalue). We see that in regions in which the spread of
$\ket{g(\tilde{q})}$ with respect to $\ket{g(q)}$ is big
(quantified in terms of the variance of $D(\omega)$), in other
words in regions where two ground states differ significantly, the
overlap  will be small. Recently, Quan {\em et al.}
\cite{zanardi-china} established a link between the critical
behavior of the environment and quantum decoherence, showing that
the Loschmidt echo \cite{Loschmidt,Loschmidt1}  $L(q,t)$ of the environment
exponentially goes to zero as we approach the regions of
criticality. A simple algebra gives us that the Fourier transform
of the projected density of states is precisely the Loschmidt
echo, $|\int_{-\infty}^{+\infty}D(\omega)e^{-i\omega
t}d\omega|^2=L(q,t).$
 We see that the kinematics of a system,
given by the geometry of ground states through the overlap
 function, influences its dynamics as well:
the smaller the overlap i.e., broader $D(\omega)$ the faster
the decay of the Loschmidt echo.

\section{Conclusions and Discussion} In this paper, by discussing
the examples of the Dicke and $XY$ spin-chain models, we have
presented a characterization of quantum phase transitions
based on the study of the scaling and asymptotic behaviors of the
overlap between two ground states taken in two
close points of the parameter space. Though this quantity might in general
not provide an efficient numerical tool it is conceptually quite appealing.
In fact the ground state overlap is a purely Hilbert-space geometrical quantity, whose investigation
does not rely on any a priori understanding of the specific kind
of order patterns or peculiar dynamical correlations hidden in the analyzed system.
While in the case of the Dicke Hamiltonian it was possible to
analyze the overlap function directly in the thermodynamical
limit, the case of the $XY$ model is more subtle.
The process of ``orthogonalization" between two ground states has
two physically different mechanisms in the case when two states
belong to two different phases: one is a common decrease of the
overlap due to infinite number of sub-systems in thermodynamical
limit, the other is characteristic for the case of QPT and is due
to different structures of ground states in different quantum
phases.



\section{acknowledgments.}
NP is funded by the European Commission, contract No.
IST-2001-39215 TOPQIP. The authors greatly acknowledge the
discussions with A. T. Rezakhani.




\begin{thebibliography}{}

\bibitem{sachdev} S.~Sachdev, {\em Quantum Phase Transitions}, Cambridge University Press (1999).



\bibitem{qis} For reviews, see A. Steane, Rep. Prog. Phys. {\bf 61}, 117
(1998); D. P. DiVincenzo and C. H. Bennett, Nature {\bf 404}, 247 (2000).



\bibitem{osterloh} A.~Osterloh, L.~Amico, G.~Falci, and R.~Fazio, Nature 416, 608 (2002).
\bibitem{vidal} G.~Vidal, J.I.~Latorre, E.~Rico, and A.~Kitaev, \prl {\bf 90}, 227902 (2003).


\bibitem{zanardi} Y.~Chen, P.~Zanardi, Z.D.~Wang, and F.C.~Zhang, New J. Phys. (2006) quant-ph/0407228 .
\bibitem{others}S.-J~Gu, G.-S.~Tian, and H.-Q.~Lin, quanth-ph/0509070.
\bibitem{lidar} L.-A.~Wu, M.S.~Sarandy, and D.A.~Lidar, \prl {\bf 93}, 250404 (2004).

\bibitem{carollo} A.~Carollo, and J.K.~Pachos, Phys. Rev. Lett. {\bf 95}, 157203 (2005); Shi-Liang
Zhu, Phys. Rev. Lett. {\bf{96}}, 077206 (2006) ; A. Hamma, quant-ph/0602091.

\bibitem{cejnar} P. Cejnar, V. Zelevinsky, and V. V. Sokolov, Phys. Rev. E {\bf 63}, 036127 (2001).

\bibitem{other} A. Volya, and V. Zelevinsky, Phys. Lett. B {\bf 574}, 27 (2003); V. Zelevinsky, B. A. Brown, N.
Fraizer, and M. Horoi, Phys. Rep. {\bf 276}, 85 (1996); P. Cejnar, and J. Jolie, Phys. Rev. E {\bf 58}, 387
(1998); P. Cejnar, and J. Jolie, Phys. Rev. E {\bf 61}, 6237 (2000).




\bibitem{anderson} P.W. Anderson, Phys. Rev. Lett. {\bf 18}, 1049
(1967).


\bibitem{zanardi-china}  H.T. Quan, Z. Song, X.F. Liu, P. Zanardi and C.P.
Sun,  Phys. Rev. Lett.
{\bf{96}}, 140604 (2006)




\bibitem{emary} C. Emary and T. Brandes, Phys. Rev. E, 066203
(2003).

\bibitem{primakoff} T. Holstein and H. Primakoff, Phys. Rev. {\bf
58} 1098 (1949).


\bibitem{ITP} J. von Neumann, Comp. Math. {\bf 6}, 1 (1938); T Thiemann and O Winkler, Class. Quantum Grav. {\bf 18}
4997 (2001).


\bibitem{sachdev-catastrophe} S. Sachdev., M. Troyer and M. Vojta, Phys. Rev. Lett. {\bf 86}, 2617 (2001)

\bibitem{Loschmidt}  P. R. Levstein, G. Usaj and H. M. Pastawski, J. Chem. Phys. 108, 2718 (1998); G. Usaj, H. M. Pastawski P. R. Levstein, Mol. Phys.95, 1229 (1998);
H. M. Pastawski, P. R. Levstein, G. Usaj, J. Raya and J. Hirschinger, Physica A 283, 166 (2000)

\bibitem{Loschmidt1}  Z. P. Karkuszewski, C. Jarzynski and W. H.
Zurek, Phys. Rev. Lett. {\bf 89}, 170405 (2002);  F.M. Cucchietti,
D.A.R. Dalvit, J.P. Paz and W.H. Zurek, Phys. Rev. Lett. {\bf 91},
210403 (2003).




\end{thebibliography}
\end{document}